\documentclass[sigconf]{acmart}
\AtBeginDocument{%
  }

\setcopyright{acmlicensed}
\copyrightyear{2025}
\acmYear{2025}
\setcopyright{acmlicensed}\acmConference[WWW Companion '25]{Companion Proceedings of the ACM Web Conference 2025}{April 28-May 2, 2025}{Sydney, NSW, Australia}
\acmBooktitle{Companion Proceedings of the ACM Web Conference 2025 (WWW Companion '25), April 28-May 2, 2025, Sydney, NSW, Australia}
\acmDOI{10.1145/3701716.3715217}
\acmISBN{979-8-4007-1331-6/2025/04}

\usepackage{balance}
\usepackage{caption}
\usepackage{subcaption}
\usepackage{soul}
\usepackage[inline]{enumitem}
\usepackage[normalem]{ulem}
\usepackage{multirow}
\usepackage{graphicx} 
\usepackage{adjustbox} 
\usepackage{array}
\usepackage{algorithm}
\usepackage{algpseudocode}
\usepackage{amsmath}
\usepackage{makecell}
\usepackage{colortbl}
\usepackage{amsmath}

\begin{document}

\newcommand{\E}{\mathbf{E}}
\newcommand{\x}{\mathbf{x}}
\newcommand{\y}{\mathbf{y}}
\title{Dimension Mask Layer: Optimizing Embedding Efficiency for Scalable ID-based Models}

\author{Srijan Saket$^*$}
\email{srijanskt@gmail.com}
\orcid{0006-9460-1203}
\affiliation{
    \institution{ShareChat}
    \city{Seattle}
    \country{USA}
}

\author{Ikuhiro Ihara$^*$}
\email{tsukue@gmail.com}
\orcid{0003-5299-3788}
\affiliation{
    \institution{ShareChat}
    \city{New York}
    \country{USA}
}

\author{Vaibhav Sharma}
\email{sharmavaibhav1729@gmail.com}
\orcid{0000-8376-8979}
\affiliation{
    \institution{ShareChat}
    \city{Bangalore}
    \country{India}
}

\author{Danish Kalim}
\email{danishkalim14@gmail.com}
\orcid{0004-0944-262X}
\affiliation{
    \institution{ShareChat}
    \city{Bangalore}
    \country{India}
}

\renewcommand{\shortauthors}{Srijan Saket, Ikuhiro Ihara, Vaibhav Sharma, and Danish Kalim}

\begin{abstract}
  In modern recommendation systems and social media platforms like Meta, TikTok, and Instagram, large-scale ID-based features often require embedding tables that consume significant memory. Managing these embedding sizes can be challenging, leading to bulky models that are harder to deploy and maintain. In this paper, we introduce a method to automatically determine the optimal embedding size for ID features, significantly reducing the model size while maintaining performance.

 Our approach involves defining a custom Keras layer called the dimension mask layer, which sits directly after the embedding lookup. This layer trims the embedding vector by allowing only the first N dimensions to pass through. By doing this, we can reduce the input feature dimension by more than half with minimal or no loss in model performance metrics. This reduction helps cut down the memory footprint of the model and lowers the risk of overfitting due to multicollinearity.

 Through offline experiments on public datasets and an online A/B test on a real production dataset, we demonstrate that using a dimension mask layer can shrink the effective embedding dimension by 40-50\%, leading to substantial improvements in memory efficiency. This method provides a scalable solution for platforms dealing with a high volume of ID features, optimizing both resource usage and model performance.
\end{abstract}




\maketitle

\def\thefootnote{*}\footnotetext{Equal Contribution}

\section{Introduction}
In recent years, using embeddings to represent large sets of categories or IDs has become a key part of machine learning models, especially in areas like recommendation systems, natural language processing (NLP), and computer vision \cite{pan2019social, almeida2019word, srijanmonitoring}. To put the scale into perspective, approximately 3.7 billion videos are uploaded daily to YouTube and 23 million to TikTok \cite{scaleYT, scaleTikTok}. Music streaming platforms like Spotify offer a library of 100 million songs \cite{scaleSpotify} for users to choose from. Learning meaningful representations for IDs at such a massive scale poses production challenges. Embeddings map categorical variables or words into dense \(n\)-dimensional vectors, making it easier for models to detect patterns from input data. However, determining the optimal size for these embeddings is challenging and often based on guesswork. Too few dimensions can limit the model’s ability to capture important information, while too many can inflate the embedding table size, leading to larger memory footprints. This can result in inefficient training and decreased accuracy due to issues like multicollinearity.

\begin{figure}[!t]
\begin{minipage}[b]{\linewidth}
    \includegraphics[width=\textwidth]{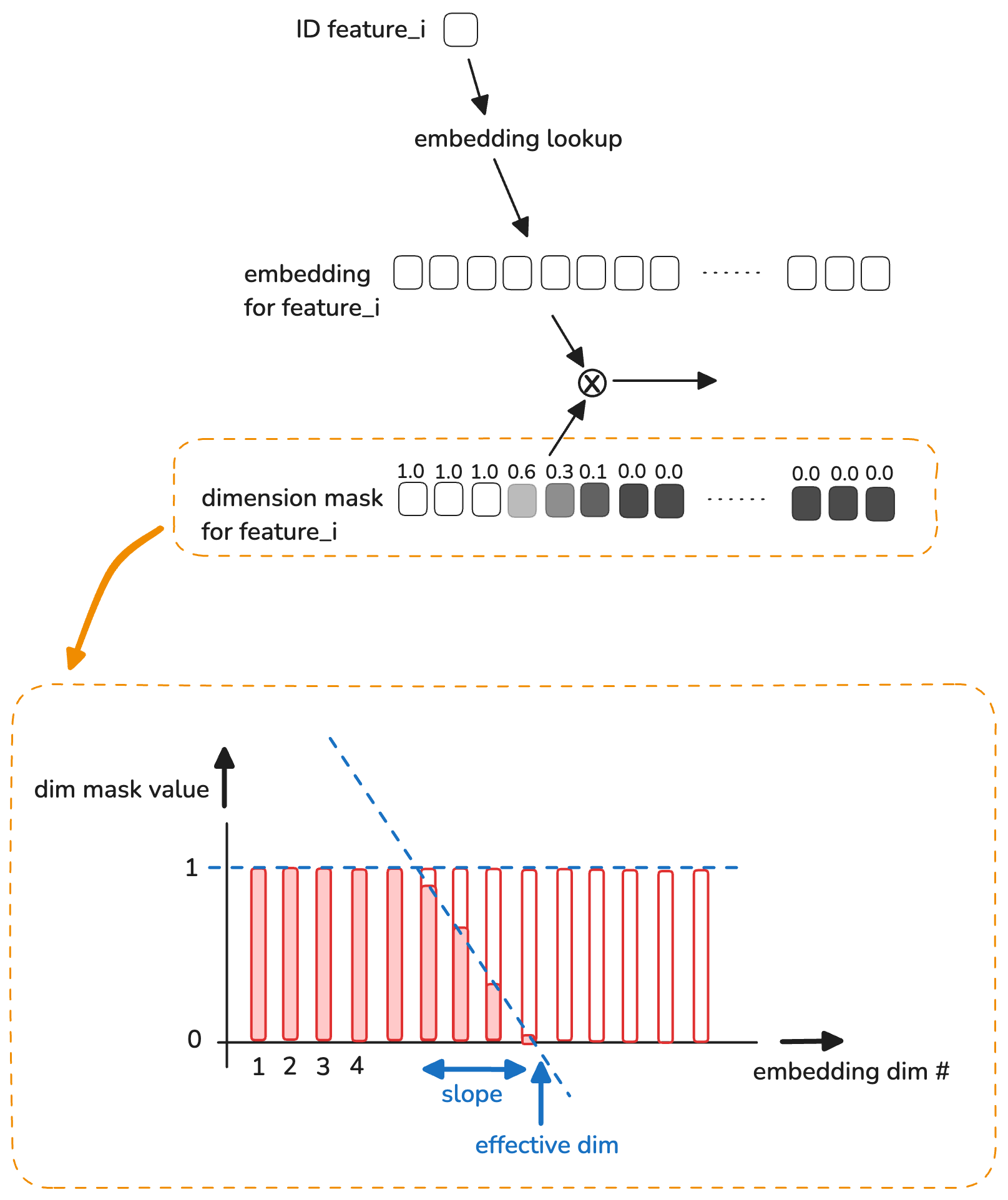}
\end{minipage}
\caption{The diagram above illustrates how the dimension mask layer only lets the first N (effective\_{dim} in the diagram) dimensions through, (with some smooth gradation at the end)}
\label{fig:dml}
\Description{DML working}
\end{figure}

This paper introduces the \textbf{Dimension Mask Layer (DML)}, a novel approach that adjusts the size of the embeddings automatically during training. Instead of manually deciding the size, DML lets the model find the right size on its own, improving both efficiency and performance. By hiding unnecessary dimensions, DML ensures only the most important information moves through to deeper layers of the network, reducing model complexity and boosting the model’s ability to generalize well.

There are two ways in which embedding sizes can influence model performance and  cost. The first is through the total size of the model’s input layer. Reducing the total embedding dimensions decreases the model's memory footprint. For example, if the feature layer has \(K\) nodes that are fully connected to the next layer with \(L\) input nodes, the model must maintain \(K \times L\) learnable weights. Since both \(K\) and \(L\) tend to be large numbers, keeping \(K\) is crucial for overall model efficiency. Removal of redundant inputs also helps reducing the chance of multicollinearity, a problem that can arise when embedding sizes are unnecessarily high, leading to overfitting.

The second factor is the total size of the model’s embedding hash table, which is a function of both embedding sizes and the cardinalities of ID features. In our production setup, for instance, the embedding size of the userID feature alone dictates the model’s binary size. Reducing the embedding size of high-cardinality features, such as userID, is essential for keeping the model size manageable. Although this implementation of DML does not directly address this second aspect, as feature cardinalities are not factored into its optimization, one can still utilize DML by first mnanually constraining the sizes of critical features, and then letting DML to optimize the remaiing embedding sizes, as demonstrated in Section~\ref{exp_sharechat_dataset}

\section{Related Work}
Optimizing the size of embeddings has been a tough problem across different fields including recommendation system, natural language processing and foundational models. Earlier methods for word embeddings like Word2Vec and GloVe \cite{church2017word2vec, pennington2014glove} used fixed sizes, and researchers had to rely on trial and error to find the right fit. 

Techniques, like Principal Component Analysis (PCA) \cite{abdi2010principal} and embedding quantization, have been suggested to handle the issue of embedding size. For instance, recent advancements in embedding quantization \cite{sbertEmbeddingQuantization, shakir2024quantization, microsoftReduceVector} have shown how reducing embeddings into smaller chunks can help save memory. However, these methods typically compress the embeddings after training, while DML adjusts the size on the go during training, making it more responsive to the model’s needs.

Ginart et al. \cite{ginart2021mixed} proposed a method to dynamically adjust embedding dimensions based on the activity of individual items. While this approach offers flexibility, it introduces challenges for industrial systems where real-time training and change in item activity are inevitable, making it hard to maintain consistency in production environments. Shi et al. \cite{shi2020compositional} introduced a double hashing technique that combines quotient and remainder based hashing to reduce memory usage while avoiding collisions. Final embeddings is derived by combining representations from two separate embedding tables, using operations such as concatenation, addition, or element-wise multiplication. Although effective, this method requires extensive experimentation to fine-tune parameters, such as the hash function base for each of the different features, to balance memory savings and performance trade-offs. 

The Matryoshka Representation Learning (MRL) model \cite{kusupati2022matryoshka} also has some common ground with our approach. MRL builds layered embeddings that gradually get more detailed, learning from rough to fine patterns in a shared structure. Though this is good for learning at different levels, MRL doesn’t adjust the embedding size while training—it sticks to preset sizes. In contrast, the DML adapts the size based on what the model learns during training.

A recent study from TensorFlow on training recommendation systems with dynamic embeddings, TensorFlow Recommender Addons (TFRA), talks about managing size of the embedding table by removing less relevant item embeddings \cite{tensorflowTrainingRecommendation}. It does so by using a feature called \textit{restrict()}, which gets triggered based on certain user-defined policies. While both TFRA and DML approaches optimize memory and performance, the former addresses table size, and the latter deals with adjusting embedding dimensionality during training. 

Newer research shows that the optimal size for embeddings doesn’t just depend on the number of categories but also on how many tasks the model needs to handle and the hidden patterns in the data. For example, GateNet \cite{huang2020gatenet} introduced a way to manage how much information moves through the layers of a network, which is somewhat similar to how we mask unnecessary dimensions. While GateNet focuses on routing data, the core idea aligns with our goal of filtering out irrelevant dimensions. Another interesting approach called Plug-in Embedding Pruning (PEP) \cite{liu2021learnable} prunes embeddings by introducing learnable thresholds for each parameter. During training, these thresholds identify and remove less important parameters by setting them to zero. In contrast, DML adjusts the number of active dimensions for the entire embedding vector, masking unnecessary dimensions as a whole. PEP uses learnable thresholds to enforce sparsity whereas DML relies on a regularization term along with a learnable masking mechanism.

To sum it up, while existing solutions deal with embedding size by using fixed approaches or adjusting them after training, DML offers a flexible, real-time way to adjust embedding dimensions as the model trains. This can lead to better accuracy, less memory use, and removes the need to manually tune embedding sizes.

\section{Dimension Mask Layer Algorithm}
\subsection{Basic Concept} \label{basicConcept}
The Dimension Mask Layer (DML) is placed immediately after the embedding lookup during initial training. As illustrated in Figure~\ref{fig:dml}, it allows only the first \textit{effective\_dim} dimensions of the underlying embedding vector to pass through, masking the remaining dimensions. The model's loss function encourages an increase in \textit{effective\_dim} to leverage more dimensions, while a regularization term encourages to minimize \textit{effective\_dim}. This helps the model strike a balance between  efficiency and performance, using only as many dimensions as necessary. 

The detailed process is  described in Algorithm~\ref{algorithm}. If there are multiple embeddings in a model, all embedding sizes can be tuned simultaneously by wrapping each embedding with a new instance of DML.

\begin{figure}[!t]
\begin{minipage}[b]{\linewidth}
    \includegraphics[width=\textwidth]{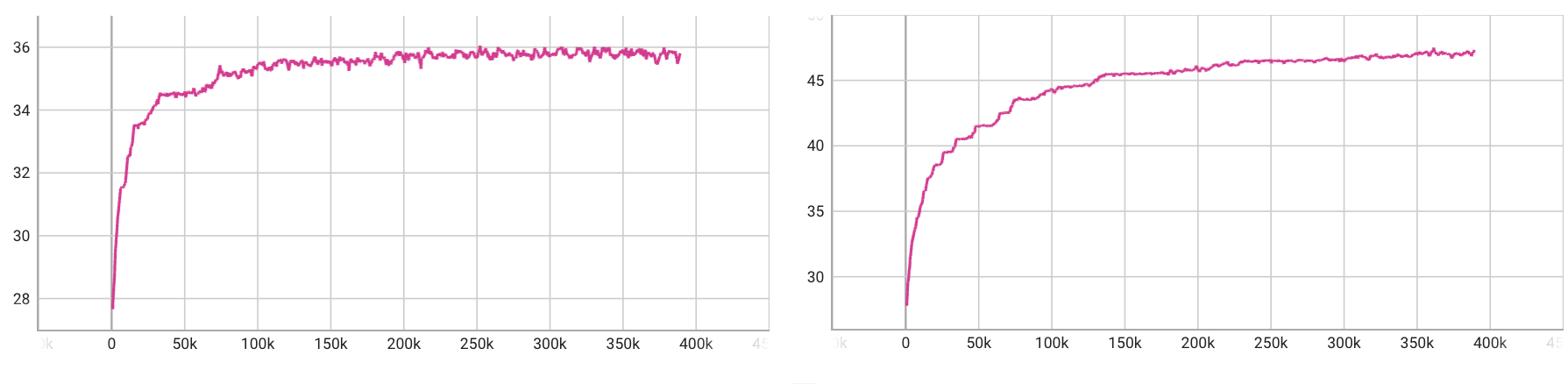}
\end{minipage}
\caption{Example output of effective\_dim}
\Description{}
\label{fig:tensorboard}
\end{figure}

Figure~\ref{fig:tensorboard} is an example output of Tensorboard, where each \textit{effective\_dim} converges to an optimal value as the training progresses. At the end of the training run, the DML can be removed, and replaced with the hard-coded dimension size determined by the DML.

\subsection{Implementation Details}

\begin{algorithm}
\caption{Dimension Mask Layer Algorithm}
\label{algorithm}
\begin{algorithmic}[1]
\Require $\text{input\_shape} \in \mathbb{R}^{B \times d}$: input shape, where B is the batch size and the d is the original dimension of the input embedding.
\Require $\text{initial\_effective\_dim} \in [0, d]$ : initial value of the $\text{effective\_dim}$, where d is the original dimension of the input embedding.
\Require $\text{slope} \in (0, \infty)$: hyper parameter with a default value of 2.0.
\Require $\text{alpha} \in (0, \infty)$: hyper parameter with a default value of 5.0.
\Require $\text{regularizer}$: regularization method with a default of $\text{L1}(0.001)$
\Statex
\State \textbf{Initialize} DimensionMaskLayer class with $\text{regularizer}$
\Statex
\Procedure{build}{$\text{input\_shape}$}
    \State $\text{original\_dim} \gets \text{input\_shape}[1]$
    \State $\text{scaled\_effective\_dim} \gets$ single trainable variable initialized with the value ($\text{initial\_effective\_dim} / \text{input\_shape}$), regularized by the specified regularizer.
\EndProcedure
\Statex
\Procedure{call}{$\mathbf{inputs}$}
    \State $x_{2} \gets \max(0, \text{scaled\_effective\_dim} * \text{original\_dim})$
    \State $x_{1} \gets x_{2} - \text{slope}$
    \State $\mathbf{x} \gets [0, 1, \dots, (\text{original\_dim} - 1)]$
    \State $\mathbf{mask} \gets \frac{1}{x_{1} - x_{2}} * \mathbf{x} + \frac{x_{2}}{x_{2} - x_{1}}$
    \State $\mathbf{mask} \gets \max(0, \min(1, \mathbf{mask}))$
    \State log the value of $x_{2}$ as the current optimal dimension.
    \State \Return $\text{apply\_mask}(\mathbf{inputs}, \mathbf{mask})$
\EndProcedure
\Statex
\Procedure{apply\_mask}{$\mathbf{inputs}, \mathbf{mask}$}
    \State $\mathbf{pseudo\_random\_dropout} \gets \text{sigmoid}(\text{alpha} * (2.0 * \mathbf{mask} - \text{random.uniform}(\mathbf{input\_shape}) - 0.5))$
    \State \Return $ \mathbf{inputs} \times \mathbf{pseudo\_random\_dropout}$
\EndProcedure
\end{algorithmic}
\end{algorithm}

\subsubsection{Dynamic Control of \textit{effective\_dim}}
In Algorithm~\ref{algorithm},  \textit{effective\_dim} ($\in [0,d]$) is a continuous decimal value, which makes the loss function smooth and differentiable with respect to \textit{effective\_dim}.

The total model loss is the sum of model's accuracy loss and the regularization loss, where the scaled \textit{effective\_dim} ($\in [0,1]$) from all DMLs in the model are incorporated into the regularization term in the form of $ L1 $ or $ L2 $. (Equatiion~\ref{eq:regLoss})

\begin{equation} \label{eq:regLoss}
\mathcal{L}_{\text{regularization}} = \sum_{\scriptsize \text{all instances of DML}}\| \mathit{scaled\_effective\_dim} \| 
\end{equation}

\begin{equation}
\mathcal{L}_{\text{total}} = \mathcal{L}_{\text{accuracy}} + \mathcal{L}_{\text{regularization}}
\end{equation}

This incentives the model to minimize \textit{effective\_dim}, unless the performance improvement from the use of more dimensions outweighs regularization penalty.

The final value of $\lceil \textit{effective\_dim} \rceil$ at the end of the training is considered the optimal embedding dimension.  
 
\subsubsection{Application of the \textit{mask} value} \label{applicationOfMask}
There is a \textit{mask} value ($\in [0,1]$) for each corresponding dimension of the underlying embedding. This value determines how each dimension of the raw embedding vector is allowed to pass through.

A simplest application would be to multiply the raw embedding vector by the \textit{mask} vector, which empirically works in most cases. But this allows the model to cancel out the effect of the \textit{mask} by simply adjusting the weights of connections to downstream nodes. To counter this, DML uses a pseudo-gate function using random values combined with a sigmoid function, similar to dropout\cite{srivastava2014dropout}.More details in Section~\ref{section:conversionOfMaskValue}


\subsubsection{Hyperparameters}
Apart from the selection of the regularizer and its weight, there are two additional hyperparameters in Algorithm~\ref{algorithm}.
\begin{itemize}
\item \textit{slope}: Determines how many adjacent dimensions are ramped-up simultaneously. A higher value lets more dimensions to be exposed to back propagation from early on, enabling a smoother transition of  \textit{effective\_dim}. The recommended default value is 2.0, meaning that one additional dimension is gradually warmed up alongside its neighboring dimension. See Figure~\ref{fig:dml} for more details.
\item \textit{alpha}: Controls the sensitivity on how the \textit{mask} value ($\in [0,1]$) is translated into a random dropout probability. The recommended default is 5.0. Further details are provided in the later Section~\ref{section:alpha}
\end{itemize}

\subsubsection{Conversion from \textit{mask} value(\( x \)) to the Pseudo Random Dropout Multiplier(\( z \))}
\label{section:conversionOfMaskValue}
As described earlier in Section~\ref{applicationOfMask}, the underlying embedding value is not simply multiplied by the \textit{mask} value. It will instead be multiplied by a pseudo random dropout multiplier \( z \) which is derived as follows:

Let \( x \in [0, 1] \) be the \textit{mask} value and \( y \sim \mathcal{U}(0, 1) \) an uniformly distributed random variable. The transformed multiplier \( z \) is defined as:

\begin{equation}\label{multiplier}
     z = \frac{1}{1 + \exp\left(- \alpha \times (2x - y - 0.5)\right)}
\end{equation}

This is a sigmoid function with stochastic behavior added by the random variable \( y \). For a smaller value of \( x \), \( z \) is distributed around lower value ranges near 0, stochastically suppressing the impact of the underlying embedding to the downstream neural network, while not completely blocking it. As \( x \) increases, the distribution of \( z \)  shifts toward a higher value range closer to 1, effectively passing through the embedding value most of the time.

\begin{figure}[!t]
\begin{minipage}[b]{\linewidth}
    \includegraphics[width=\textwidth]{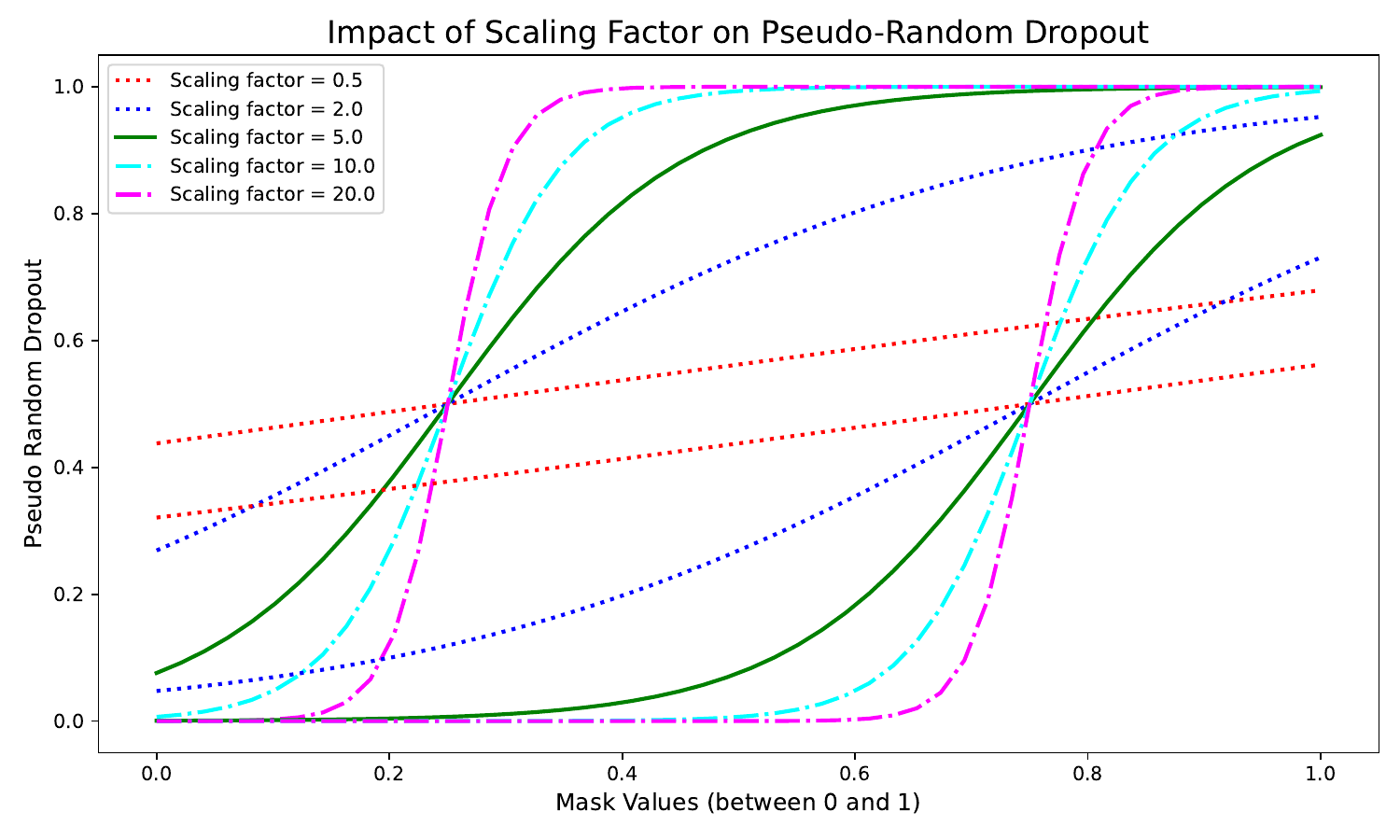}
\end{minipage}
\caption{Impact of $\alpha$ on the behavior of Eq \ref{multiplier}.}
\label{fig:scaling_factor}
\end{figure}

\subsubsection{Impact of the Hyperparameter $\alpha$}
\label{section:alpha}
Figure~\ref{fig:scaling_factor} shows how the choice of $\alpha$ affects the sensitivity of the gate function.

High $\alpha$ values, such as 10.0 and 20.0 shown in Figure~\ref{fig:scaling_factor}, result in the gate being fully closed (value of z pinned at 0.0) when the \textit{mask} value is low, and fully open when \textit{mask} value is high. This leads to a flat gradient with respect to the \textit{mask} (and subsequently, to \textit{effective\_dim}) in these regions, making optimization impossible.

On the other hand, low $\alpha$ values, such as 0.5 or 2.0 reduces the model’s sensitivity to different \textit{mask} values, which can result in a slower convergence.

\section{Experiments and Observations}
We have conducted two experiments for analysis and observations presented in this paper.
\begin{itemize}
    \item The first experiment is in ShareChat ads production environment, where we evaluated the result of DML on production traffic, with the intention of improving cost efficiency without sacrificing accuracy. 
    \item The second experiment is on the open-source Avazu dataset, which we use to demonstrate the general utility of DML and to test the effect of different parameter configurations.
\end{itemize} 

\subsection{Experiment with ShareChat Production Data} \label{exp_sharechat_dataset}
\subsubsection{Data Context}

\begin{table}[h!]
    \centering
    \resizebox{\columnwidth}{!}{%
    \renewcommand{\arraystretch}{1.8} 
    \begin{tabular}{|>{\centering\arraybackslash}p{0.15\textwidth}|>{\centering\arraybackslash}p{0.4\textwidth}|}
    \hline
    \large \textbf{Feature Category} & \large \textbf{Description} \\
    \hline
    \textbf{User Features} & 
    \textbf{Demographic features}: Age, gender, and hashed geographic location. Sampled for uniform distribution.\newline
    \textbf{Content preference embeddings}: Based on user engagement with non-ad content on ShareChat.\newline
    \textbf{App affinity embeddings}: Based on past apps installed by the user on the platform. \\
    \hline
    \textbf{Ad Features} & 
    \textbf{Ad categorical features}: Characteristics such as ad size and category.\newline
    \textbf{Ad embedding}: Represents the video or image content of the ad. \\
    \hline
    \textbf{Interaction Features} & 
    \textbf{Count features}: User interactions with ads, advertisers, and categories across time windows. \\
    \hline
    \textbf{Label} & Click or Install \\
    \hline
    \textbf{Data Structure} & Each row has a unique ID representing an ad impression, noting if it led to a click or install. \\
    \hline
    \end{tabular}%
    }
    \caption{Summary of ShareChat Data Features}
    \label{tab:sc_dataset}
\end{table}

We used ad impression and engagement data from multiple product surfaces, described in  \cite{saket2024crafting}. We only considered impressed views, meaning recommendations that were actually seen by the users. We did not apply any sampling for unbiasedness, but note that the data may be biased by the model’s own ranking. The data has 500M–800M data points per day, which translates to approximately $\approx 9.1 \text{B}$ data points in total. The feature details are summarised in Table~\ref{tab:sc_dataset}.

\subsubsection{Baseline}
The baseline production model is a MTL MMoE model which uses a fixed embedding size of 48 for most ID features, with some exceptions of low cardinality features having smaller sizes. Notably, the embedding size for userID is set to 8. Given its high cardinality (on the order of \textit{millions}), keeping this particular embedding small is essential for maintaining a manageable model size in production. The sum of all embedding dimensions is 2431, as shown in the first column of Table~\ref{tab:dim_reduction_results}

\subsubsection{Experiment Setup}
We follow a two-step experimental setup: an offline test followed by an online test.

First, we prepared the treatment model, which is the exact copy of the baseline model but with DML applied to all embeddings except for userID. The embedding size for userID was fixed at 8 for the reasons discussed earlier. The optimal embedding sizes were selected by training a model on 14 days of production data ($\approx 9 \text{B}$ data points), 

We then compared the offline metrics (details in Section~\ref{eval_metrics}) on one day of validation data. The results, as well as hyper parameter configurations and the total embedding dimension count determined by DML, are shown in the second column of Table~\ref{tab:dim_reduction_results}. Additionally, we trained a model with userID embedding size fixed at 4, and another model with userID  wrapped by DML, allowing it to also determine the optimal size of userID embedding. The results are shown in the third and forth columns of Table~\ref{tab:dim_reduction_results}, although these two configurations were only for reference and not taken forward to the online test.

In the online setup, both models were evaluated in a budget-aware A/B test \cite{liu2020trustworthy}. In a budget-aware A/B test, the ad campaign budget is split equally, with each portion allocated to a randomly selected user group. This ensures that the groups are independent and do not compete for budget allocation.

We implemented the configuration found in the offline experiment above (as in the second column of Table~\ref{tab:dim_reduction_results}), and compared its online performance against the baseline model over a 14-day period, with both models updated daily via incremental training.

\definecolor{LightCyan}{rgb}{0.88,1,1}
\newcolumntype{b}{>{\columncolor{LightCyan}}c}

\begin{table}[h!]
\centering
\resizebox{\columnwidth}{!}{%
\begin{tabular}{|l|b|b|c|c|}
    \hline
    & \textbf{Baseline} & \textbf{Treatment} & \makecell{\textbf{UserID size} \\ \textbf{fixed at 4}} & \makecell{\textbf{UserID size} \\ \textbf{unconstrained}} \\
    \hline
    \textbf{UserID dim} & 8 & 8 & 4 & 43 \\
    \hline
    \textbf{Total dim count} & 2431 & 1164 & 907 & 948 \\
    \hline
    \textbf{(\% of Baseline)} &  & 47.88\% & 37.31\% & 39.00\% \\
    \hline
    \textbf{Hyper parameters} & &  \multicolumn{3}{|c|}{ \makecell{$ initial\_effective\_dim=3, slope=2.0,$ \\ $regularizer\_weight=0.0, \alpha=5.0 $} } \\
    \hline
    \textbf{Click AUC} & 0.8042 & 0.8038 & 0.8037 & 0.8040 \\
    \hline
    \textbf{Click RCE} & 10.66 & 10.65 & 10.63 & 10.63 \\
    \hline
    \textbf{Install AUC} & 0.9158 & 0.9395 & 0.9173 & 0.9176 \\
    \hline
    \textbf{Install RCE} & 16.57 & 16.70 & 16.47 & 16.37 \\
    \hline
\end{tabular}%
}
\caption{Model Comparison across Different UserID Dimension Settings}
\label{tab:dim_reduction_results}
\end{table}

\subsubsection{Evaluation Metrics}\label{eval_metrics}
Two metrics, Relative Cross Entropy (RCE) \cite{belli2020privacy} and Area Under the Curve (AUC), were used for the offline evaluation.

RCE measures the performance of a prediction model as percentage improvement from the naive baseline model, and defined as:

\begin{equation}
(\text{CE}_{\text{baseline}} - \text{CE}_{\text{pred}}) \times 100 / \text{CE}_{\text{baseline}}
\end{equation}

Where \( \text{CE}_{\text{baseline}} \) is the average cross entropy (CE) of the naive baseline model, and  \( \text{CE}_{\text{pred}} \) is the average CE of the model to be evaluated. AUC is a
commonly used evaluation metric primarily focused on ranking, while RCE takes both ranking and calibration into account.

In the online setup, we also measured key business metrics such as click through rate, conversion rate, revenue, and user retention.

\subsubsection{Training and Inference Environment Details} 
Both models were trained using 6 workers, each with 6 CPUs, 48 GB memory, and an NVIDIA TESLA T4 accelerator. Three parameter servers (4 CPUs, 48 GB memory) were used.

\subsubsection{Key Observations and Results}
We have successfully reduced the total embedding dimensions by more than 50\% without significant loss in both online and offline metrics.

\begin{figure}[!t]
\begin{minipage}[b]{\linewidth}
    \includegraphics[width=\textwidth]{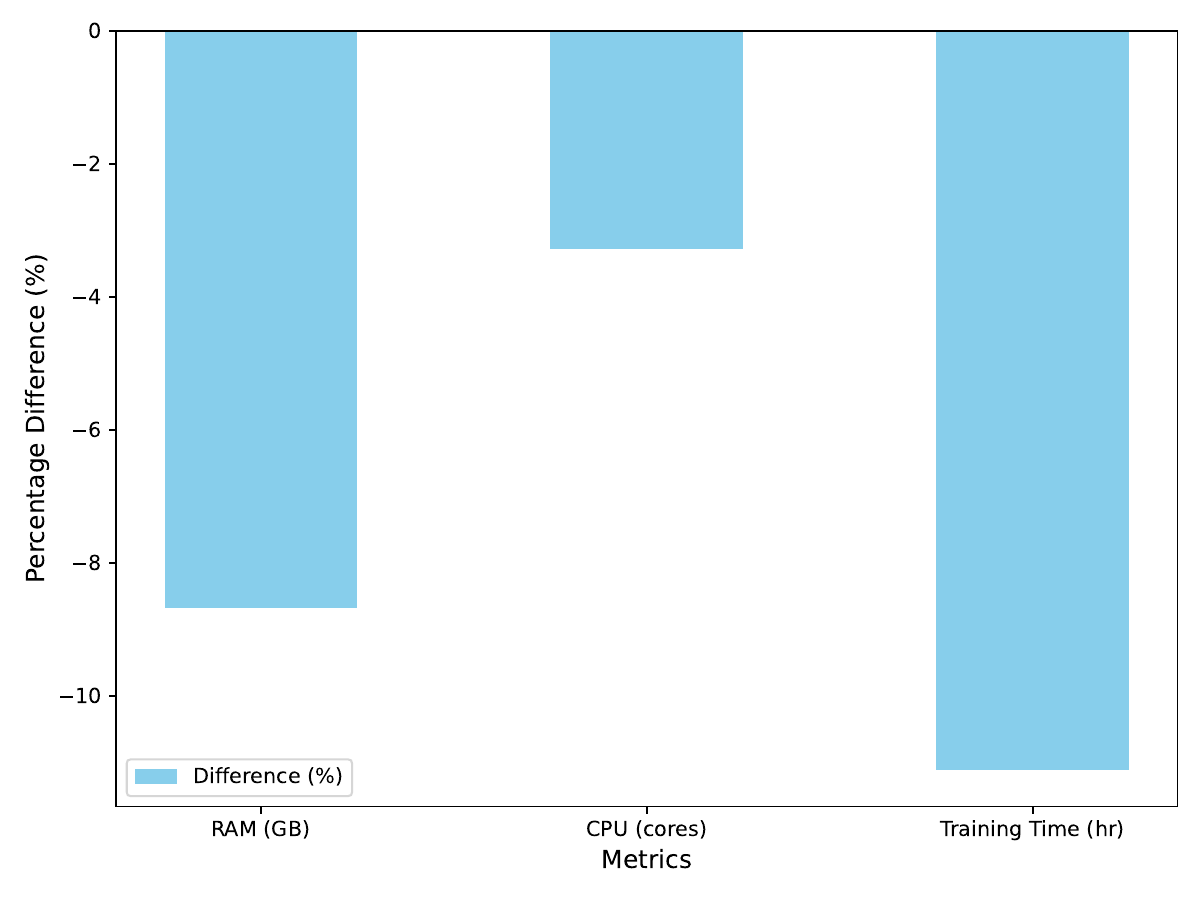}
\end{minipage}
\caption{Relative Percentage Difference Between Baseline and DML Across Key Resource Metrics}
\Description{Comparison of the performance of the DML Model relative to the Baseline across various key metrics}
\label{fig:sc_online_metrics}
\end{figure}

Additionally, the new model achieved lower online inference costs, with average memory usage reduced by $8.7\%$ and average CPU usage by $3.3\%$. Average training time also decreased by \textbf{$11.1\%$}, as shown in Figure~\ref{fig:sc_online_metrics}. However, the overall cost impact is somewhat limited due to: 1. minimum deployment requirements in production systems, which results in a fixed cost during non-peak hours, and 2. the dimension size only affecting the first few layers of the model, with no change to the computational graph in subsequent layers.

One notable observation is that initializing with a minimum \textit{effective\_dim} and a very low regularizer weight (or even without  regularization) and letting DML expand \textit{effective\_dim} empirically worked better than starting with a maximum \textit{effective\_dim} and letting DML reduce it. This may be explained by the fact that this setup lets DML incrementally add dimensions one at a time, until additional dimensions no longer provide utility, at which point DML naturally stops increasing \textit{effective\_dim} any further.

\begin{figure*}[t!]
    \centering
    \begin{subfigure}[t]{0.45\textwidth}
        \centering
        \includegraphics[width=\textwidth]{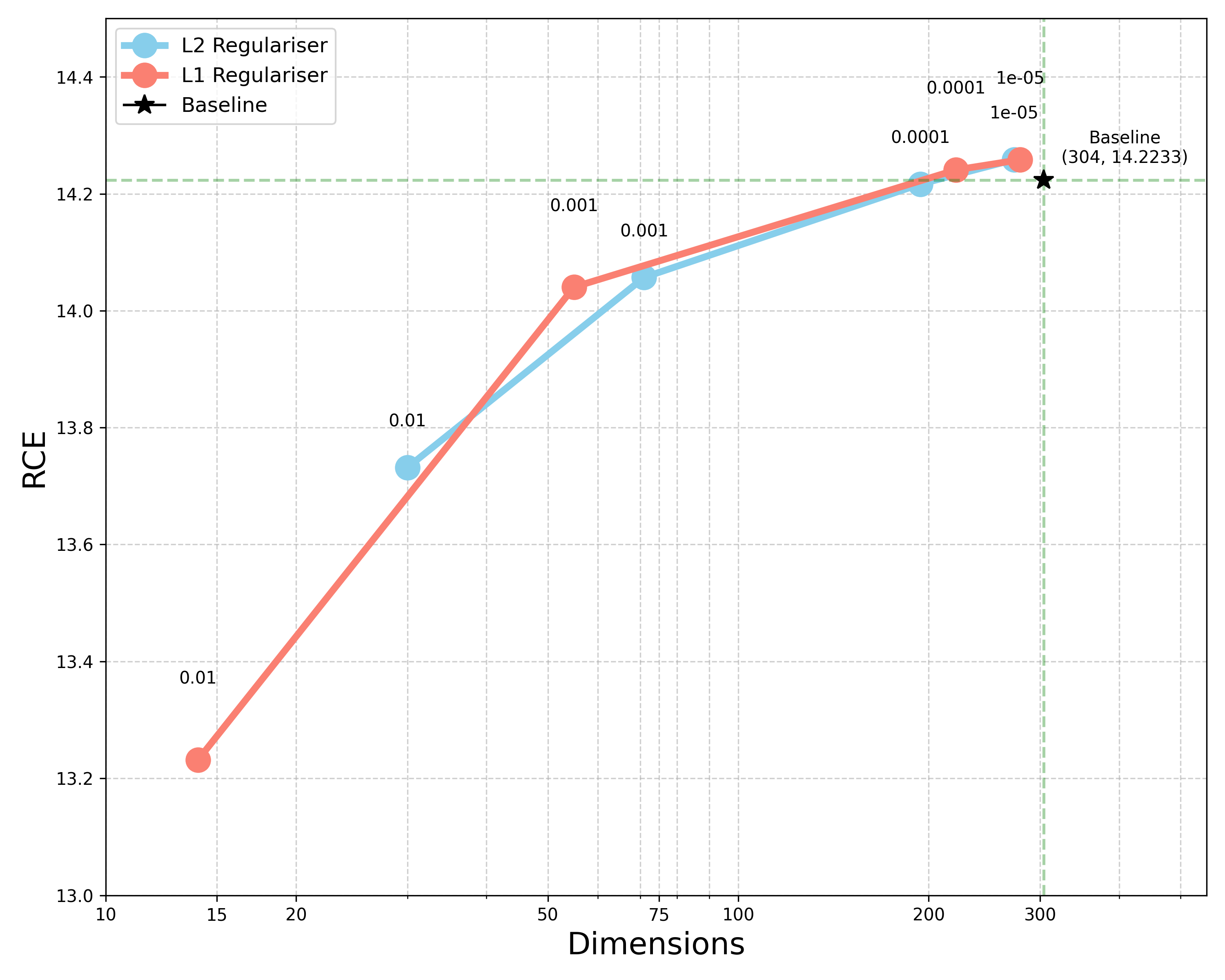}
        \caption{Effect of DML for various choices of regulariser and weight}
        \label{fig:regulariser}
    \end{subfigure}
    \hspace{1mm}
    \begin{subfigure}[t]{0.45\textwidth}
        \centering
        \includegraphics[width=\textwidth]{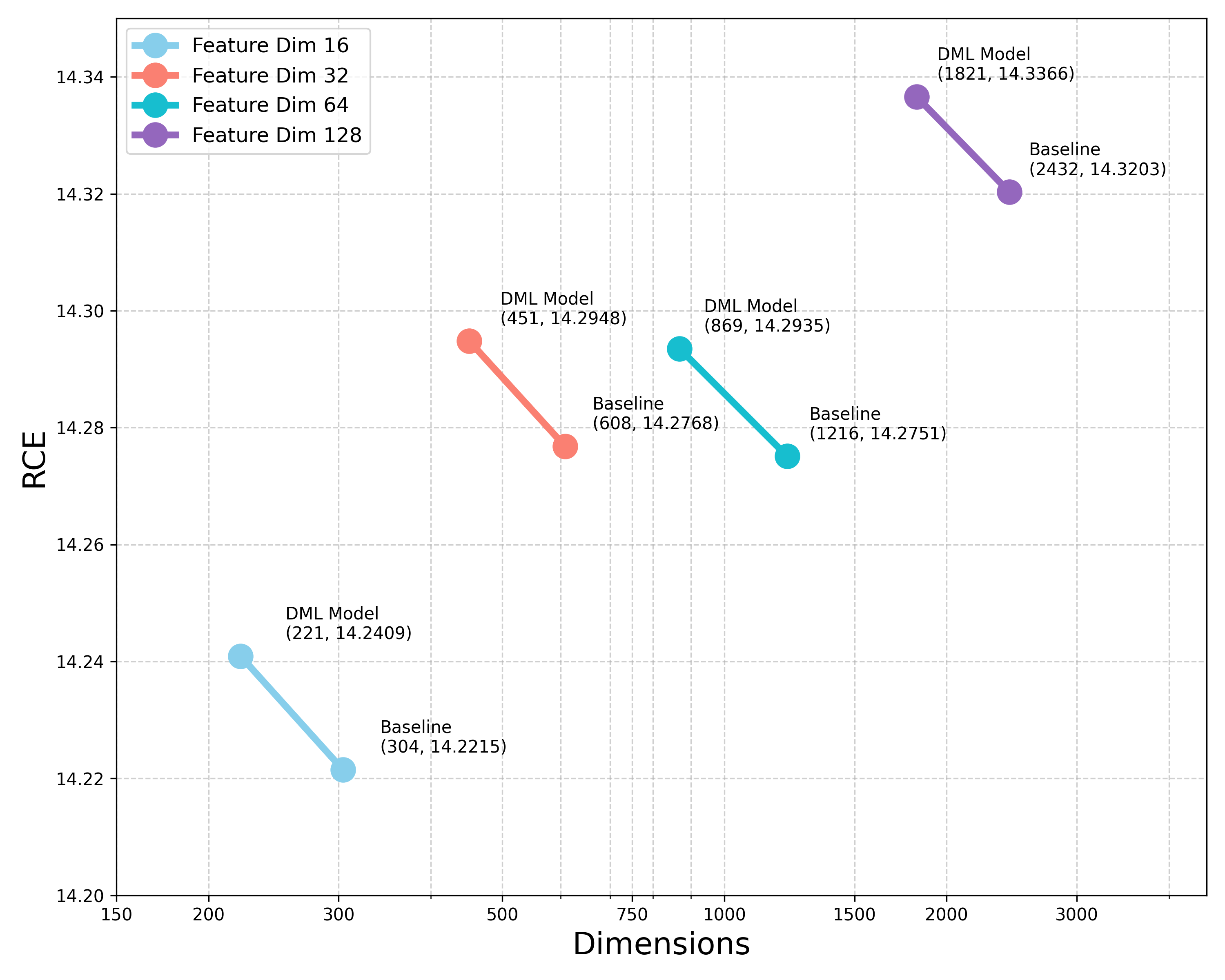}
        \caption{ 
        Effect of DML for various baseline embedding dims
        }
        \label{fig:baselinedims}
    \end{subfigure}
    \caption{Key Observations on Avazu Dataset with Different Parameter Settings}
    \vspace{-2mm}
\label{fig:avazu}
\Description{...}
\end{figure*}

\subsection{Experiments with Avazu Dataset} \label{avazuExp}
\subsubsection{Data Context}
Avazu is a public dataset \cite{avazu-ctr-prediction} with 10 days of click-through data for training and 1 day of data for testing. It contains over 40 million rows of ad click data, with features such as id, click (indicating if an ad was clicked), hour (timestamp), device attributes (like device type and connection type), and site or app details. The features are mostly categorical and describe the context in which an ad was displayed, making this dataset well-suited for applying DML.

\subsubsection{Evaluation Metrics}
Here, we compare RCE and the total embedding dimension size to measure trade-offs between performance and efficiency. We selected RCE as a representative metric for model performance, as we found it to be the most reliable in our past experiments.

\subsubsection{Experiment Setup}
In each experiment below, we trained multiple models with DML, under varying configurations, comparing the results against the baseline model without DML.
\begin{enumerate}
    \item Compare the effect of using different regularizers \{$ L1, L2 $\} and regularizer weights \{$ 1\mathrm{e}{-2}, 1\mathrm{e}{-3}, 1\mathrm{e}{-4}, 1\mathrm{e}{-5} $\}, with a fixed baseline embedding size of 16.
     \item Evaluate the effect of DML on different baseline embedding dimensions \{$ 16, 32, 64, 128 $\}, with regularizer L1, weight $ 1\mathrm{e}{-4} $.
 \end{enumerate}

Other hyperparameters were fixed to
\{ $  \mathit{initial\_effective\_dim}=14, 
 slope=2.0, \alpha=5.0 $ \} for the treatment model.

\subsubsection{Implementation Details}
We used a straightforward architecture with two linear layers of 128 and 64 nodes, each followed by ReLU activation. A classification head with sigmoid activation was applied on top of these layers. Training was conducted on a single NVIDIA Tesla T4 GPU. Each categorical feature in the Avazu dataset was bucketised and passed through an embedding lookup. These embeddings were then concatenated to form the input to the aforementioned architecture. For the DML model, we simply wrapped the output of the embedding lookup with DML.

\begin{figure}[!t]
\begin{minipage}[b]{\linewidth}
    \includegraphics[width=\textwidth]{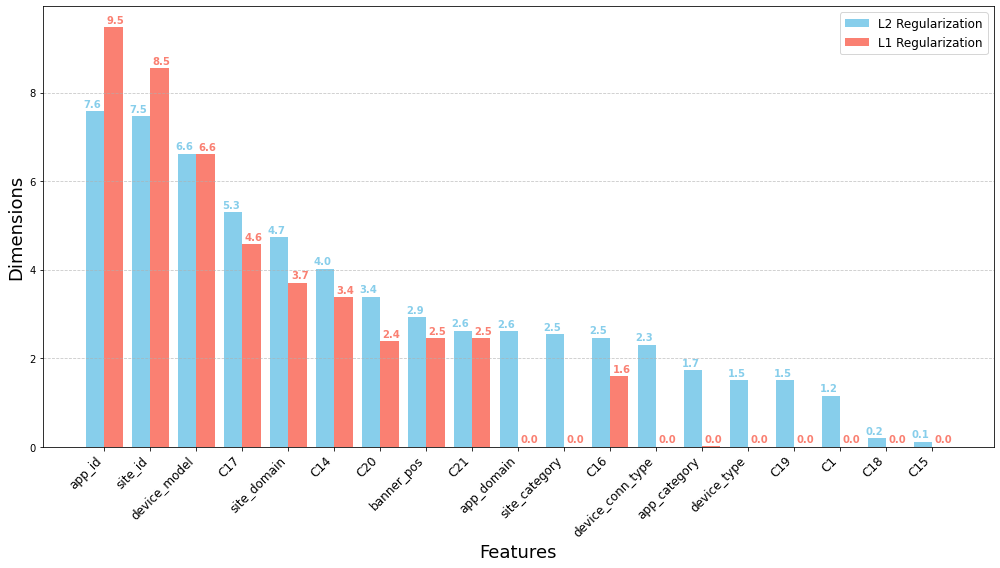}
\end{minipage}
\caption{Effect of L1 and L2 Regularizers, weight=$ 1\mathrm{e}{-3}$}
\Description{}
\label{fig:L1vsL2}
\end{figure}

\subsubsection{Key Observations and Results}
Figure~\ref{fig:regulariser} shows how the choice of the regularizer weight affects the balance between the model's accuracy and the total embedding size. Higher the weight, smaller the resulting model would be, while sacrificing more accuracy. Use of the L1 regularizer tends to result in a higher number of feature with 0 dimensions compared to L2. Figure~\ref{fig:L1vsL2} shows the difference in resulting dimensions between L1 and L2 regularizers, both with weight=1e-3.

We should also emphasize that with the minimum regularizer weight, DML was able to find a configuration that beats the baseline model in both accuracy and efficiency (top right corner in the graph).

Figure~\ref{fig:baselinedims} shows how the impact of DML changes for different initial baseline embedding sizes. We see here too that DML finds a configuration that outperforms the baseline in both model size and accuracy in all cases.

\section{Conclusion}

We introduced the Dimension Mask Layer (DML), a method which automatically finds optimal embeddings sizes in a single training run (Section~\ref{basicConcept}). This approach has proven effective in a production setting (Section~\ref{exp_sharechat_dataset}), and is also adaptable to other general usecases (Section~\ref{avazuExp}).

Through carefully crafted offline and online experiments, we have also shown that DML not only achieves both higher accuracy and higher efficiency than the baseline model, but can also be tuned to achieve a tailored balance between accuracy and efficiency suited for specific demands (Section~\ref{avazuExp}).

\section{Future Work}
One potential direction for future work is to extend the Dimension Mask Layer (DML) to serve as a feature importance tool. This can be achieved by applying DML to dense features, treating each feature as a one-dimensional embedding. In this setup, a feature can be considered more important if the corresponding DML's \textit{effective\_dim} approaches closer to 1, or converges faster toward 1.0. On the other hand, \textit{effective\_dim} approaching closer to 0 would be a sign that the feature is less important. Although we explored this approach with promising results, it is beyond the scope of this paper. One key insight from this experiment though, is the importance of initializing \textit{effective\_dim} to its maximum value and allowing the model some training time before DML is allowed to start reducing the dimensions. This adjustment is crucial in situations where original size of the embedding is very small. In such cases, \textit{slope} parameter becomes ineffective, resulting in the dimension reduction happening too quickly, causing the model to underestimate feature significance.

Another area of improvement is making DML more feature-aware by incorporating feature-specific attributes such as cardinality or platform importance. These attributes could be baked into the loss function to improve the effectiveness of dimension reduction. Additionally, understanding the trade-offs between reducing the dimension of a feature and considering its cardinality or importance is crucial. Balancing these factors could lead to more efficient models by selectively reducing the complexity of less important features while preserving the value of key ones.

\bibliographystyle{ACM-Reference-Format}
 \balance
\bibliography{sample-base}

\end{document}